# Waste Animal Bone-derived Calcium Phosphate Particles with High Solar Reflectance

*Nathaniel LeCompte[1], Andrew Caratenuto[1,2], and Yi Zheng[1,2]\**

[1]Planck Energies, Brighton, MA 02135, United States

[2]Department of Mechanical and Industrial Engineering, Northeastern University, Boston, MA 02115, United States

ABSTRACT

Highly reflective Calcium Phosphate (CAP) nanoparticles have been obtained from waste chicken and porcine bones. Chicken and pork bones have been processed and calcined at temperatures between 600ºC and 1200ºC to remove organic material and resulting in CAP bio-ceramic compounds with high reflectance. The reflectivity of the materials in the solar wavelength region is on par with chemically synthesized CAP. The high reflectivity, consistently over 90%, as well as the size distribution and packing density of the nanoparticles obtained in these early bone studies make a strong case for pursuing this avenue to obtain pigment for high solar reflectivity applications, such as passive daytime radiative cooling. The results presented indicate a viable path toward a cost-effective and eco-friendly source of highly reflective cooling pigments. By sourcing calcium phosphates from animal bones, there is also the potential to divert large quantities of bone waste generated by the meat industry from landfills, further contributing toward sustainability and energy reduction efforts in the construction industry and beyond.

**Introduction**

Climate change caused by the combustion of carbon-based fuels and ongoing environmental damage being caused by the production of forever chemicals are all but certainly the most pressing issues of our time. To combat climate change and reduce energy consumption for cooling purposes, highly reflective, bio-compatible pigments for cooling applications are being developed in earnest (1, 2). The potential of this type of pigment for roof paints and coatings as well as for other cooling applications is substantial. This is particularly pertinent as the world is navigating the mounting climate crisis and works to mitigate the effects of rising temperatures and reduce energy consumption to escape the worst possible outcomes for our climate and Earth (3). Highly reflective calcium phosphate particles synthesized via wet-chemistry precipitation have proven to be effective as pigments for cooling coatings such as cool roof paints (4). Cost-effective sources of CAP are being explored to improve the sustainability and associated production costs of such pigments with a focus on bio-based and biowaste sources (5, 6). To this end, a comparable CAP pigment has been developed from waste animal bone. Chicken and pork bones from chicken wings and spiralized ham respectively were collected, cleaned and processed into fine powders (7, 8, 9). This powder as well as the processing methods were studied and compared to previously demonstrated CAP cooling pigment, currently synthesized from laboratory grade chemicals (4, 8). Fabricating the cooling pigment from bones that would otherwise be thrown away would not only make CAP production eco-friendlier but contribute to greenhouse gas reduction efforts more broadly (11). With negligible cost, and the added benefit of diverting large amounts of animal bone waste from landfills, CAP upcycled from chicken, pork and other animal bones is seen as a highly desirable alternative to synthesized CAP for use in cooling pigments and paints currently under development (12, 13). A wide range of preliminary testing has been completed to obtain CAP from both chicken and pork bones, using various processing and calcination methods.

**Materials and Methods**

Approximately two dozen unique processes have been conducted thus far. Raw chicken wing and pork bones were obtained from food service suppliers. To clean the bones and remove organic tissue, the bones were immersed in boiling water for around one hour (14, 15). The bones were then manually cleaned by hand and with a brush in a lab sink to remove any visible tissue that remained. A portion of the bones also went through a process of deproteinization by immersion in a NaOH solution (7). The clean bones were subsequently placed in a dryer at 60°C until dried completely (12 – 72 hours). Following this, the bones were either calcined directly to fully remove any residual organic matter, then ground in an electric grinder, or ground in a ball mill or electric grinder before calcination to obtain the CAP powder. The obtained bones and bone powder were calcined at a range of temperatures between 600ºC and 1200ºC (9). At all studied temperatures, whole bones and bone powders were calcined for 2 hours, with calcination being performed either once or twice. The calcined CAP powders were then added to a powder sample holder before reflectance over the solar spectrum (200nm to 2500nm) was measured using a Jasco V770 spectrometer. The data was subsequently normalized to obtain a single indicative reflectance value for comparison. Because the bone-derived CAP particles were unable to be pressed into a puck using a cylindrical mold, to measure the bulk material directly, as was done with synthesized CAP, the powders were placed in a powder sample holder to obtain the reflectance measurements. Experiments conducted with synthesized CAP powders showed that bulk samples return reflectance values 4% to 7% higher than the same samples measured in the powder sample holder. Due to this, it is estimated that the materials' true reflectance is on average 5.87% higher than the recorded measurements. These values were subsequently compared with the reflection data of CAP obtained via the mentioned wet chemistry precipitation method of synthesis (4). These methods have been repeated with pork bones, achieving similar results.

An investigation into the effect of the bone marrow within the bones and its impact on reflectance was also performed. Bone samples marked as MR had their marrow removed prior to calcination. This was done by snapping the chicken bones and using small metal files to manually remove the marrow. All pork samples had their marrow removed similarly prior to calcination. Multiple processing methods were experimented with in an effort to determine the most effective CAP acquisition technique. Bones were either calcined whole after cleaning or processed and ground prior to calcination. As the internal bone marrow was expected to affect the reflectance of the ground powder, a set of chicken bones were snapped at the ends such that the marrow could be removed manually with small metal files. The remaining bone shells were the calcined whole or ground prior to calcination. To limit hands-on processing time, some bones were also calcined or ground after snapping the bone ends but without removing the marrow inside. In this case the marrow was fully exposed to oxygen so it could adequately burn off during calcination. Here, the differences between pre-ground and post-ground bones, relative to calcination, are primarily explored.

**Results**

A wide range of preliminary testing has been completed with both chicken and pork bones, varying the cleaning and calcination processes. A total of two dozen unique processes have been conducted thus far. The reflective performance of the bone powders, found in Table 1 is similar to that of CAP powder obtained for chemically synthesized by a wet chemistry precipitation method. With negligible cost, and the added benefit of diverting large amounts of animal bone waste from landfills, CAP upcycled from chicken and pork bones is seen as a highly desirable alternative to chemically synthesized CAP for use in novel cooling pigments and PDRC paints currently under development.

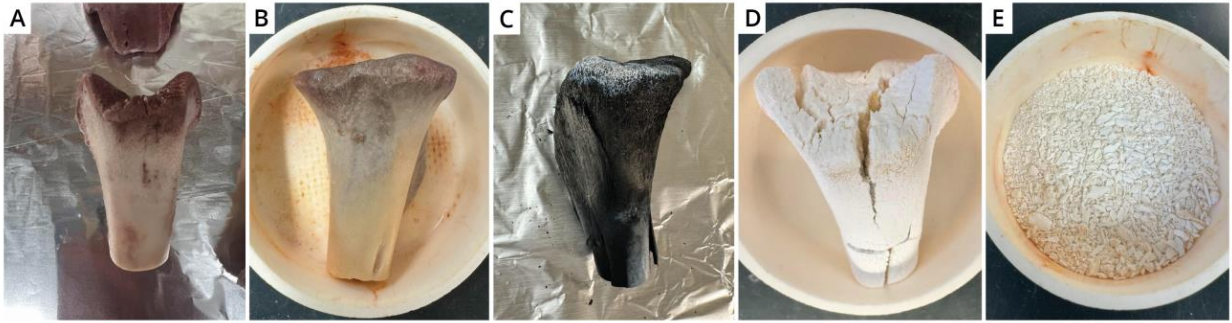

Figure 1. Pork bone at various stages of processing and calcination; A. Cleaned bone prior to deproteinization, B. Cleaned bone after deproteinization, C. Charred bone after first calcination, D. Pork bone following a second calcination at 1000ºC, E. Pork bone powder ground following second calcination.

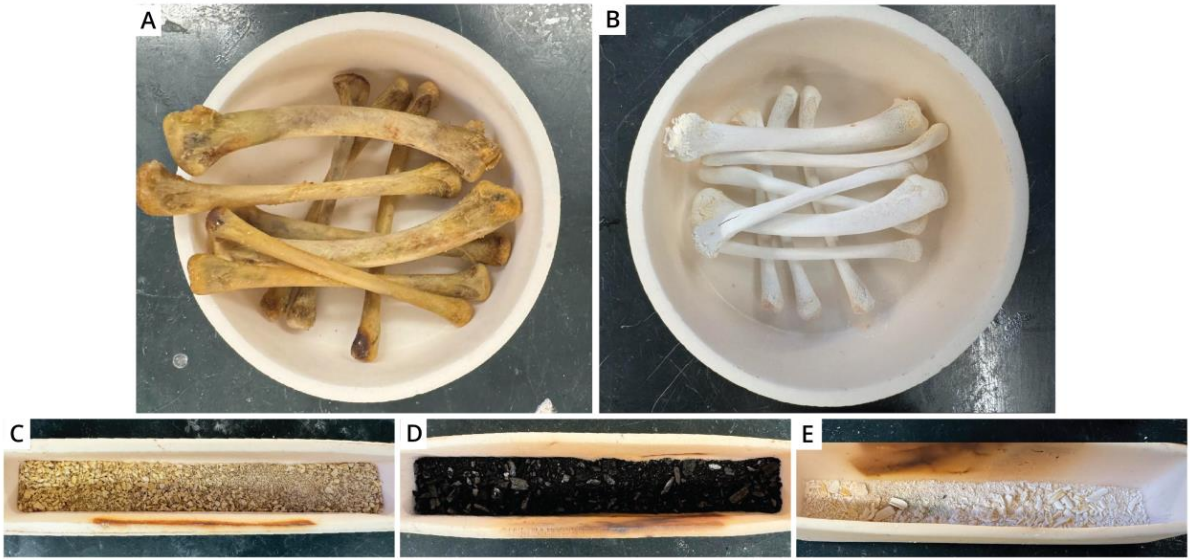

Figure 2. A. Cleaned dried chicken bones, B. Whole chicken bones calcined at 1000ºC, C. Ground raw chicken bones, D. Chicken bones after calcination at 600ºC, E. Chicken bones after second calcination at 1000ºC.

Whole chicken bones calcined at 600ºC became charred due to remaining organic matter. Once ground and calcined again at 1000ºC, the carbonized powder became very white and exhibited high reflective properties. All bones calcined under 800ºC appeared black and necessitated further calcination at higher temperatures. This was performed at 800ºC and 1000ºC.

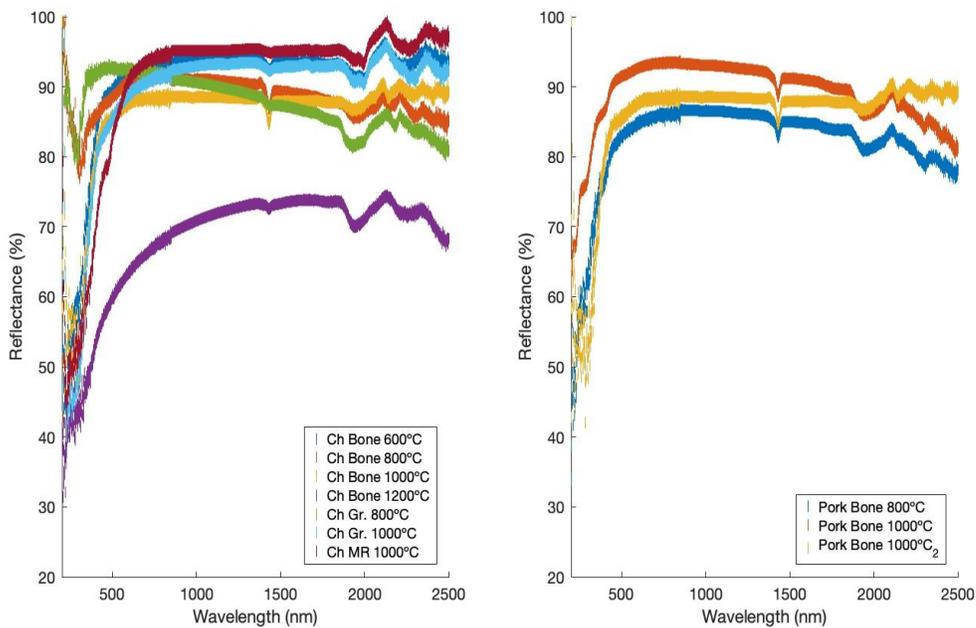

Figure 3. Reflectivity of chicken and pork bone samples across the solar spectrum; A. Reflectivity of chicken bones calcined in both whole and ground form at various temperatures, B. Reflectivity of pork bones calcined whole at 800ºC and 1000ºC.

Table 1. Trials and Results of 10 Chicken & Pork Bone Samples Calcined at 600ºC to 1200ºC

| Trial | Source | Calcination Form | 1st Calcination Temp (ºC) | Result | 2nd Calcination Temp. (ºC) | Result | Normalized Reflectance |
|---|---|---|---|---|---|---|---|
| Ch1 | Chicken | Whole | 600 | Black | 1000 | White | 90.02% |
| Ch2 | Chicken | Whole | 800 | Brown | 1000 | White | 88.92% |
| Ch3 | Chicken | Whole | 1000 | White | N/A | N/A | 80.46% |
| Ch4 | Chicken | Whole | 1200 | White | N/A | N/A | 64.82% |
| Ch5 | Chicken | Ground | 800 | White | N/A | N/A | 89.92% |
| Ch6 | Chicken | Ground | 1000 | White | N/A | N/A | 88.51% |
| Ch7 | Chicken | Whole MR | 1000 | White | N/A | N/A | 84.42% |
| P1 | Pork | Whole | 800 | Black | 800 | White | 84.14% |
| P2 | Pork | Whole | 800 | Black | 1000 | White | 90.87% |
| P3 | Pork | Whole | 800 | Black | 1000 | White | 85.68% |

The lowest calcination temperature tested with chicken bones was 600ºC. The bones turned black and resembled charcoal on their surface. These bones were ground (Fig. 2D) and calcined again at 1000° C. The resulting bone powder was no longer black and mostly white with only tinges of brown remaining. Calcination of chicken bones at 800ºC left the bones with brown spots after the first calcination. After a second calcination at 1000ºC, hardly any brown color remained. Chicken bones with their marrow removed came out white following calcination. Interestingly, the trial with the lowest reflectance was whole chicken bone calcined at 1200ºC. This is likely due to HAP and CAP compounds converting to B-TCP under higher calcination temperatures; a phenomenon that occurs over temperatures of 1000ºC. Chicken bones ground prior to calcination, namely trials Ch1, Ch5, and Ch6, resulted in very white powders with high reflectance. Calcined whole pork bones obtained similarly high reflectance, with trial P2 resulting in the most reflective pigment measured. More research will need to be done to see which compounds dominate each of the bone samples and whether bones from different animals could reach similar or higher reflectance that is beneficial to cooling performance. Calcium phosphate compounds such as HAP and calcium pyro phosphate have been shown to exhibit better cooling performance than those such as beta-TCP which dominate samples that are calcined above 1000°C.

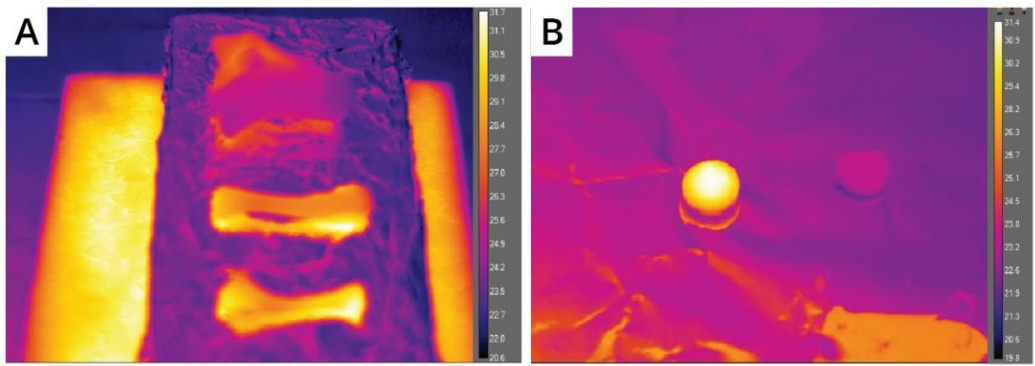

Figure 4. A. Infrared heat map of calcined pork bone (top) and chicken bones (middle & bottom) under one sun. B. An infrared heat map of pre-ground chicken bones (left) and post-ground pork bone (right) formed into pucks using a metal press following calcination at 800ºC.

Figure 4A depicts pork and chicken bones after calcination at 1000ºC. These were photographed under one sun after the bone temperatures were allowed to stabilize under the solar simulator for 10 minutes. Figure 4B indicates that ground calcium phosphates from pig bone exhibit better cooling performance than the calcium phosphates from chicken bones, pictured here on the left. A surface temperature difference of roughly 5°C to 8ºC can be seen between the chicken bone sample on the left and pork bone on the right. This sample of pork bone ground after calculation at 800°C exhibited the highest reflectance values of all bone trials and can be seen from the IR heat map to exhibit strong cooling performance under one sun. Taken together, it can be said that pork bones may be a better source of calcium phosphates for PRC applications than chicken bones, although further testing and trials with other sources of animal bones should be explored to find the best source candidate.

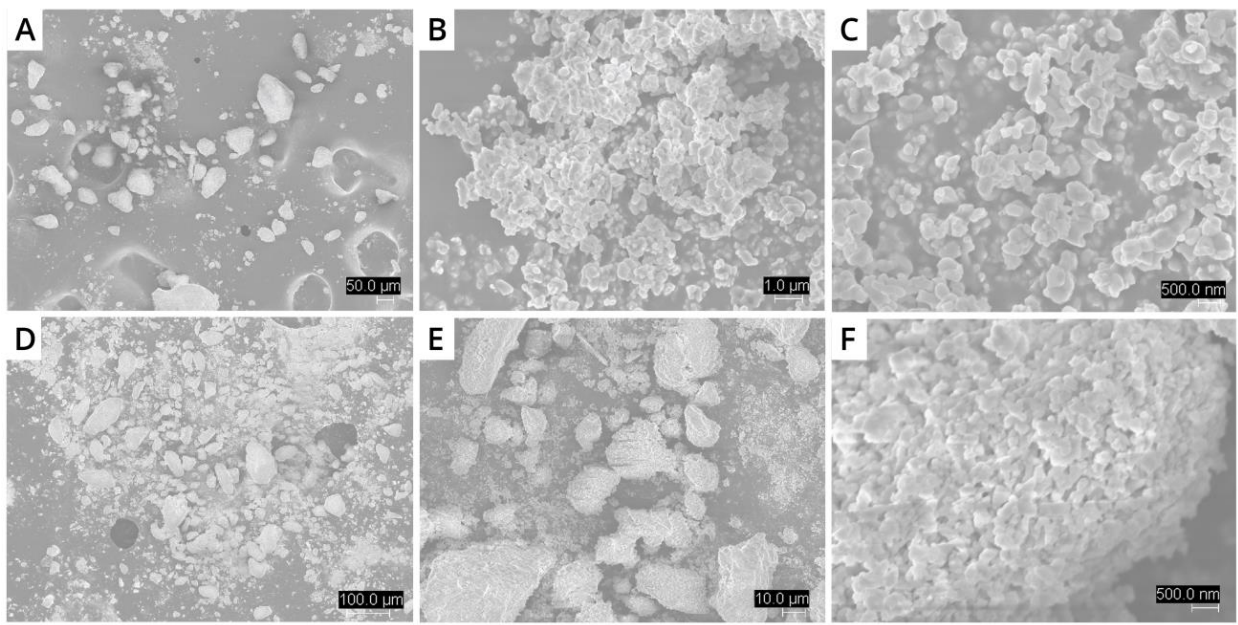

Figure 5. SEM images of ground chicken and pork bones at microscale and nanoscale. A – C depict pork bone images, and D – F depict chicken bone images.

The SEM images in Figure 4 provide valuable insight into the reflective performance of the bone-derived CAP. The distribution in particle sizes at both the micro and nano scale indicate high reflectance over the solar wavelength spectrum. The looser packing density of the pork bone particles, particularly at nanoscale is another indication of this source's better reflection performance over the ground chicken bone.

**Conclusions**

Waste bone from the food industry has the potential to be a valuable low-cost source of highly reflective CAP particles that can be used as pigment in cooling applications related to energy conservation. CAP particles derived from chicken and pork bones exhibit desirable reflective properties very similar to those of CAP synthesized from expensive lab-grade chemicals for a fraction of the cost. By replacing synthesized CAP with bone-derived CAP, not only could the price of CAP cooling pigments drop dramatically, but they could also be developed more sustainably while furthermore positively impacting global sustainability efforts. This work highlights the strong potential of a novel source for use as highly reflective, ultra-white cooling pigments for use in the rapidly expanding industry of cooling paint production which aims to boost the energy-efficiency of buildings and reduce the demand for air conditioning, particularly in places already experiencing warmer temperatures year-round due to climate change. With reflectance values of over 90%, CAP obtained from both chicken and pork bones have proven to be as effective as lab-synthesized CAP pigments and perform better than state-of-the-art cooling pigments currently utilized in the paints and coatings market. The promising results reported here warrant continued efforts toward obtaining highly reflective pigments from chicken and pork bones as well as the exploration of other bone waste sources such as cattle and fish, two animals widely cultivated by the meat and fishing industries for general consumption. While reclaiming CAP from bone waste has been researched and explored for use in the biomedical field, it has not, despite its potential, been proposed as a source of high albedo cooling pigments until now. Formulation of these pigments into coatings could pave the way to highly reflective, low-cost cooling paints that could be utilized globally and contribute greatly to electricity reduction and sustainable space cooling solutions at a time when there is a dire need for both.

AUTHOR INFORMATION

Corresponding Author

Yi Zheng − Department of Mechanical and Industrial Engineering, Northeastern University, Boston, Massachusetts 02115, United States; Department of Chemical Engineering,

Northeastern University, Boston, Massachusetts 02115, United States;

orcid.org/0000-0003-4963-9684; Email: y.zheng@northeastern.edu

Andrew Caratenuto − Department of Mechanical and Industrial Engineering, Northeastern University, Boston, Massachusetts 02115, United States,

Planck Energies, Brighton, MA, 02135, United States

Nathaniel LeCompte − Planck Energies, Brighton, MA, 02135, United States

Author Contributions

The manuscript was written through contributions of all authors. All authors have given approval to the final version of the manuscript. ‡These authors contributed equally.

Nathaniel LeCompte: Conceptualization, Methodology, Investigation, Writing – Original Draft, Visualization; Andrew Caratenuto: Conceptualization, Methodology, Investigation, Visualization; Yi Zheng: Conceptualization, Writing – Review & Editing, Supervision, Funding Acquisition.

ACKNOWLEDGMENT

This project is partially supported by the National Science Foundation through grant numbers CBET-1941743 and SBIR-2321446.

ABBREVIATIONS

CAP, Calcium phosphate; HAP, Hydroxyapatite; CPP, Calcium pyrophosphate; Wh, whole; MR, marrow-removed; B-TCP, Beta-TCP;